%% file: main.tex
\documentclass[10pt,conference]{IEEEtran}

\usepackage{amsmath}
\usepackage{amssymb}
\usepackage[frozencache,cachedir=.]{minted}
\usepackage{multirow}
\usepackage[font=small]{caption}
\usepackage{subcaption}
\usepackage{hyperref}
\usepackage[usenames,dvipsnames]{xcolor}
\usepackage{footnote}
\usepackage[T1]{fontenc}

\usepackage{listings}
\usepackage[most]{tcolorbox}
\usepackage{fancyvrb}
\usepackage{changepage}
\usepackage{csvsimple}
\usepackage{booktabs}
\usepackage{array}
\usepackage{multirow}
\usepackage{enumitem}

\usepackage{algorithm}
\usepackage{algorithmicx}
\usepackage{algpseudocode}

\input{figs/traces.tex}

\makeatletter
\newcommand{\linebreakand}{%
  \end{@IEEEauthorhalign}
  \hfill\mbox{}\par
  \mbox{}\hfill\begin{@IEEEauthorhalign}
}
\makeatother

\definecolor{correct}{RGB}{186,255,181}
\definecolor{incorrect}{RGB}{255,186,186}

\newif\ifanonymous
\anonymousfalse

\newcommand{\systemname}{\textsc{Pythoness}}

\ifanonymous
    \newcommand{\redacted}[1]{[redacted]}
\else
    \newcommand{\redacted}[1]{#1}
\fi

\title{Effective LLM-Driven Code Generation with \systemname{}}

\ifanonymous
    \author{}
\else
    \author{
    \IEEEauthorblockN{Kyla H. Levin}
    \IEEEauthorblockA{University of Massachusetts Amherst\\
        Amherst, MA, USA\\
        khlevin@cs.umass.edu}
    \and
    \IEEEauthorblockN{Kyle Gwilt}
    \IEEEauthorblockA{Williams College\\
        Williamstown, MA, USA\\
        kg15@williams.edu}
    \and
    \IEEEauthorblockN{Emery D. Berger\textsuperscript{\textdagger}}
    \IEEEauthorblockA{University of Massachusetts Amherst / Amazon\\
        Amherst, MA, USA\\
        emery@cs.umass.edu}
    \and
    \linebreakand
    \IEEEauthorblockN{Stephen N. Freund}
    \IEEEauthorblockA{Williams College\\
        Williamstown, MA, USA\\
        freund@cs.williams.edu}}
\fi

\begin{document}

\maketitle

\ifanonymous
\else

\fi

\ifanonymous
\else
\begingroup\renewcommand\thefootnote{\textdagger}
\footnotetext{Work done at the University of Massachusetts Amherst.}
\endgroup
\fi

\pagestyle{plain}

\begin{abstract}
    \input{abstract}
\end{abstract}

\input{introduction}

\input{pythoness}

\input{conclusion}

\bibliographystyle{IEEEtran}
\bibliography{main}
\end{document}

%% file: figs/traces.tex
\lstnewenvironment{run}[1]{}{}

\NewTColorBox{mybox}{ O{red} m d"" !O{} }{
    enhanced,
    colframe=#1!75!black,
    colback=#1!5!white,
    fonttitle=\bfseries,#2,
    beforeafter skip=0.25em,
    top=0em,
    bottom=0em,
    left=0.5em,
    right=0.25em,
    boxrule=0.25pt
}

\newcommand{\prompt}[1]{\texttt{\textbf{(ChatDBG) #1}}}

\DefineVerbatimEnvironment{Highlighting}{Verbatim}{commandchars=\\\{\}}

\newcommand{\subparagraph}[1]{\textbf{#1}}

%% file: abstract.tex
The advent of large language models (LLMs) has paved the way for a new
era of programming tools with both significant capabilities and risks,
as the generated code lacks guarantees of correctness and reliability.
Developers using LLMs currently face the difficult task of optimizing,
integrating, and maintaining code generated by AI. We propose an embedded
domain-specific language (DSL), \systemname{}, to address those challenges.
In \systemname{}, developers program with LLMs at a higher level of
abstraction. Rather than interacting directly with generated code, developers
using \systemname{} operate at the level of behavioral specifications
when writing functions, classes, or an entire program. These specifications
can take the form of unit tests and property-based tests, which may be
expressed formally or in natural language. Guided by these specifications,
\systemname{} generates code that both passes the tests and can be continuously
checked during execution. We posit that the \systemname{} approach lets
developers harness the full potential of LLMs for code generation while
substantially mitigating their inherent risks. We describe our
current prototype implementation of \systemname{} and demonstrate
that it can successfully leverage a combination of tests and code
generation to yield higher quality code than specifications alone.

%% file: introduction.tex
\section{Introduction}

Large language models (LLMs) have significantly impacted how developers write code.
Tools like Copilot~\cite{github_copilot} present developers with plausible code
completions and alternatives while programming. Even novice programmers can use
conversational AI systems like ChatGPT~\cite{openai_chatgpt} to create entire applications
from scratch, and professional software developers are able to author code much more
rapidly than before~\cite{paradis2024doesaiimpactdevelopment}. However, LLMs do not
necessarily reduce the total time spent programming~\cite{DBLP:conf/chi/Vaithilingam0G22},
as developers redirect the time spent writing code toward guiding and verifying the LLM's
output to ensure that it meets their
requirements~\cite{DBLP:journals/pacmpl/BarkeJP23, chopra2023conversational, kalliamvakou2022research, liang2024usability}.
This process can be tedious and error-prone, consisting of repeatedly testing and evaluating
the code, reformulating the prompt to reflect the developer's observations, and regenerating
the code. Even after this iterative development process, the code may still be considered
low quality because it is: (1) too general or not generalized enough; (2) too slow or
memory-intensive; (3) not syntactically correct; or (4) unable to pass integration tests.
Thus, there are significant challenges facing developers wishing to use LLMs to assist in code generation.

\begin{figure}[t]
    \centering
    \includegraphics[width=0.9\columnwidth]{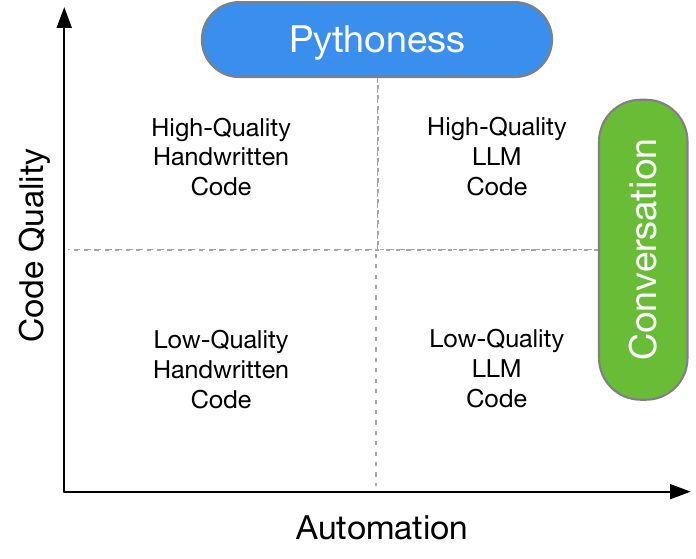}
    \caption{\label{fig:continuum}\textbf{\systemname{} lets developers integrate high quality LLM-generated code with manually written code.}
    On the automated side of the space, LLM conversations (e.g. ChatGPT) produce code with no guarantees of quality.  On the traditionally written code side,
    developers can also produce code of varying quality. \systemname{} utilizes both approaches to produce validated, high-quality code.}
    \end{figure}

\begin{table}[t]
\centering
\begin{tabular}{llll}
\toprule
\textbf{Task} & \textbf{Pre-LLM} & \textbf{Conversation} & \textbf{\systemname{}} \\
\midrule
Specify Code    & Human & Human & Human     \\
Write Code      & Human & LLM   & \systemname{} \\
Write Tests     & Human & LLM   & Human     \\
Validate Tests  & Human & Human & \systemname{} \\
Understand Code & Human & Human & \systemname{} \\
Modify Code     & Human & Human & \systemname{} \\
Fix Code        & Human & Human & \systemname{} \\
\bottomrule
\end{tabular}
\caption{\textbf{A comparison of approaches to writing and integrating LLM-generated code.}
\systemname{} abstracts away many of the
tasks associated with validating and modifying the generated code.}
\label{tab:approaches_comparison}
\end{table}

\begin{figure*}[t]
    \includegraphics[width=\textwidth]{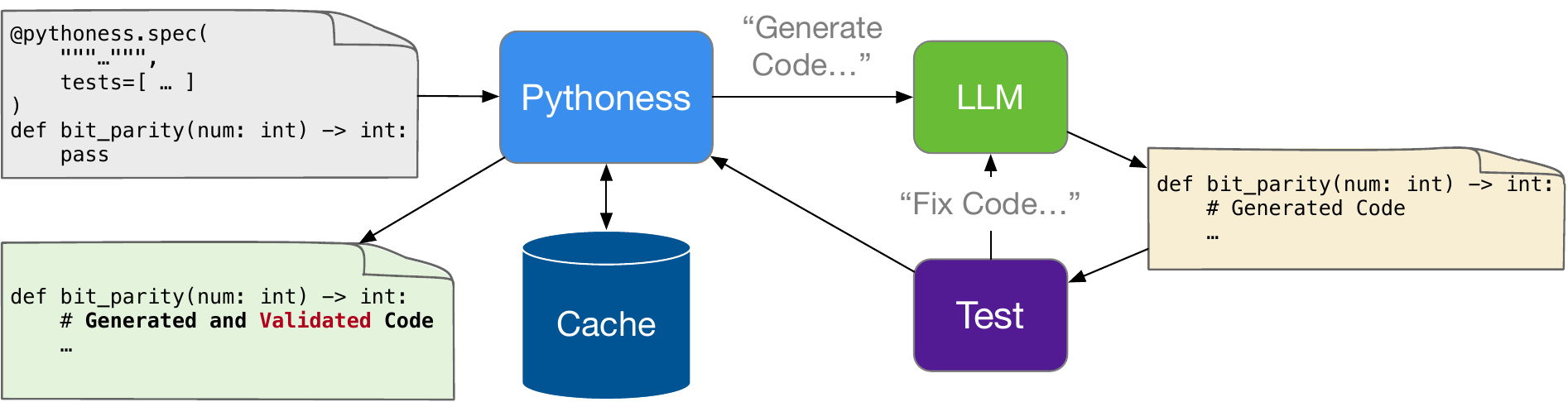}
    \caption{\label{fig:pythoness}\textbf{Overview of the software architecture of our \systemname{} prototype.}
    When a \systemname{}-decorated function is called for the first time, the \systemname{} prototype generates code
    via an LLM, checks the code against provided tests, and caches validated code in a database for future use.
    If the code fails the tests or compilation, the prototype attempts to regenerate the code until it passes all tests
    reports a failure.}
    \end{figure*}

To address these obstacles, we propose a domain-specific language called \systemname{}\footnote{
\systemname{} is named for the Oracle of Delphi from Greek mythology, a priestess said to utter
prophecies from the gods~\cite{pythia}.}. \systemname{} provides an abstraction layer above the LLM
that operates at the level of specifications and tests, guiding code generation and integrating into the 
development workflow. This Python-embedded DSL lets the developer capture function behavior through
natural language specifications, unit tests, and invariant properties of the function. From these
behavior descriptions, it generates high-quality code with built-in ``guardrails'' as run-time tests
that ensure that the code is continually validated.
Figure~\ref{fig:pythoness} presents the \systemname{} architecture.

Figure~\ref{fig:continuum} illustrates both the quality and AI-automation continuums for 
building software.
On the quality axis, human-generated code can range from poor to very well-engineered, as can the
code fully produced by AI under the right conditions. 
The \systemname{} DSL enables developers to write hybrid programs that span both approaches to
guarantee that only high-quality code is produced. 
On the manual end of the automation continuum, \systemname{} lets the developer dictate how much of
the codebase is AI-generated, whether it be only a partial function or a whole function, class, or module.
The developer can also make implementation suggestions to the LLM by providing a code template or pseudocode.
The developer can also specify structured tests against which \systemname{} will check the generated code.
\systemname{} can even grant the LLM autonomy to augment the developer's tests with additional unit
and property-based tests.

Table~\ref{tab:approaches_comparison} further highlights how \systemname{} off-loads many of the burdens of writing 
high-quality code. Similar to the conversation-driven approach, \systemname{} 
takes advantage of an LLM's ability to quickly generate code and tests. However, it also
handles code validation, modification, and maintenance. This separation of concerns leaves the human
to focus primarily or exclusively on creating code specifications and tests.

Section \ref{sec:pythoness} describes our current prototype and demonstrates its benefits
on an example from LeetCode~\cite{leetcode}. \systemname{}'s use of tests dramatically
improves the quality of generated code over using LLMs directly.
In the future, we plan to substantially expand the ways that \systemname{} validates and maintains
code quality, including through run-time and performance testing. These additional guardrails
against incorrect behavior can be run continuously and automatically, ensuring that the code remains robust and reliable.

There are many other efforts to provide guardrails for AI-generated code, including techniques
for decomposing code generation into multi-step reasoning tasks~\cite{DBLP:conf/nips/ZelikmanHPGH23} and 
for generating code that is more likely to pass co-generated tests~\cite{DBLP:conf/iclr/ChenZNZLLC23}, to name 
just a few. We believe that \systemname{} is a valuable addition to this space, as it provides a simple,
intuitive interface for developers to specify and validate code behavior.  Our framework can be readily used in conjunction with
such other techniques to further improve the robustness of the generated code.

%% file: pythoness.tex
\section{\systemname{}}
\label{sec:pythoness}

This section describes the current status of the implementation of our \systemname{} prototype.

\subsection{Initial Prototype}

The \systemname{} prototype works by letting developers specify the intended behavior of a function
via a \emph{decorator}, which guides the generation of the function; Figure~\ref{fig:headers} shows an example.
When a decorated target function is first invoked, the \systemname{} system generates the code for the function's
body using the description and tests embedded in the decoration.
The most salient aspects of the decoration header and this process are the following.

\begin{figure}[t]
    \centering
\begin{minted}{python}
@pythoness.spec(
  """Given an array nums of integers, find the
  maximum value of k for which there exist two
  adjacent subarrays of length k each in nums,
  such that both subarrays are strictly
  increasing. Assume len(nums) > 1 and consider
  a dynamic programming approach. Subarrays
  must be contiguous and non-empty sequences.""",
  tests=[
    "maxIncSubarrays([1,2,3,4,5,6])==3",
    "maxIncSubarrays([1,2,1,2,3])==2",
    "maxIncSubarrays([5,4,3,2,1])==1",
    "maxIncSubarrays([1,2])==1",
    "maxIncSubarrays([1,3,5,7,8,10,12,14,16])==4",
    "maxIncSubarrays([1,2,3,3,2,5,6,7])==2",
    "maxIncSubarrays([1,2,3,4,3,2])==2",
    "maxIncSubarrays([1,3,5,7,2,4,6,8])==4",
    "maxIncSubarrays([10,20,30,40,50,60, 100,200,300])==4",
   ],
)
def maxIncSubarrays(nums: List[int]) -> int:
  pass # Pythoness will produce code
\end{minted}
        \caption{\label{fig:headers} \textbf{The \systemname{} header used to generate the code in Figure~\ref{fig:answers}(b)}.
        For brevity, the docstring paraphrases the full \#3350 problem description on LeetCode.}
\end{figure}

\begin{figure*}[t]
    \setminted{highlightcolor=incorrect}
    \begin{tabular}[t]{cc}
    \begin{subfigure}[t]{0.48\textwidth}
\begin{minted}{python}
def maxIncSubarrays(nums: List[int]) -> int:
    n = len(nums)
    if n < 2:
        raise ValueError("The list must have more than one element.")

    inc = [1] * n

    for i in range(1, n):
        if nums[i] > nums[i - 1]:
            inc[i] = inc[i - 1] + 1

    max_k = 0
    for i in range(1, n):
        k = min(inc[i - 1], inc[i])
        if k > max_k:
            max_k = k

    return max_k

\end{minted}
        \caption{\label{fig:answer-bad}\textbf{Faulty code produced by \systemname{} when provided with the
        description in Figure~\ref{fig:headers} and not the tests.}}
    \end{subfigure}
    &
    \begin{subfigure}[t]{0.49\textwidth}
        \setminted{highlightcolor=correct}
\begin{minted}{python}
def maxIncSubarrays(nums: List[int]) -> int:
    n = len(nums)
    if n < 2:
        return 0

    left_inc = [0] * n
    right_inc = [0] * n

    for i in range(1, n):
        if nums[i] > nums[i - 1]:
            left_inc[i] = left_inc[i - 1] + 1

    for i in range(n - 2, -1, -1):
        if nums[i] < nums[i + 1]:
            right_inc[i] = right_inc[i + 1] + 1

    max_k = 0
    for i in range(n - 1):
        k = min(left_inc[i], right_inc[i + 1])
        max_k = max(max_k, k)

    return max_k + 1
\end{minted}
        \caption{\label{fig:answer-good}\textbf{Validated code produced by \systemname{} when provided with the full
        specification including unit tests in Figure~\ref{fig:headers}.}}
    \end{subfigure}
    \end{tabular}
    \caption{\label{fig:answers} \textbf{A comparison of the code produced by \systemname{} with and without unit tests.}
    Without any tests, \systemname{} produces the noticeably faulty code in Figure~\ref{fig:answer-bad} that only passes
    469 of the 1,111 tests on LeetCode. When provided with a set of unit tests, \systemname{} generates the improved code
    in Figure~\ref{fig:answer-good} that successfully passes all the LeetCode tests.}
\end{figure*}

\subsubsection{Specifying Behavior and Tests}  
A \systemname{} header includes a description of the function's purpose and expected behavior in natural language.
Additionally, the header typically includes a set of tests that the function should pass.

\systemname{} supports multiple types of tests:
\begin{enumerate}[label=(\arabic*)]
\item Simple unit tests captured as one-line assertion statements e.g. \texttt{assert f(5) == 2}.
\item Collections of unit tests defined as a \texttt{TestCase} object from the built-in Python \texttt{unittest} library. 
\item Property-based tests~\cite{DBLP:journals/sigsoft/FinkB97} describing invariants of the function's behavior
e.g. \texttt{assert fibonacci(n+2) == fibonacci(n+1) + fibonacci(n)}.
\end{enumerate}

Property-based tests are much more powerful in defining code behavior than unit tests because they represent
a property of the function being generated across a wide, possibly infinite, range of input values,
as opposed to a single input and output pair. Another key advantage of property-based tests is that
they can help prevent LLMs from overfitting to the provided unit tests.

In \systemname{}, unit tests and property-based tests can be written either as precisely-defined input values and assertions 
(as in Figure~\ref{fig:headers}) or as natural language descriptions. For example, instead of specifying
\texttt{assert f(n) \% 2 == 0}, the developer can simply write ``output of \texttt{f} is always even.''

\subsubsection{Code Generation}
\systemname{} generates code for a decorated function at run time when it is called for the first time. 
To do so, \systemname{} constructs an initial prompt to the LLM that includes the function's name and signature,
as well as the specification and tests in the \systemname{} header. It instructs the LLM to generate code that
conforms to the specification and passes the tests. \systemname{} then attempts to compile the generated code
and check its return type against the function signature. 

\subsubsection{Validation}
Once \systemname{} has a minimally-validated candidate for the function's code, it runs the function against the developer's provided tests.
For the unit tests, it directly checks that the code meets the given assertions.
For the property-based tests, \systemname{} uses the Hypothesis module~\cite{DBLP:journals/jossw/MaciverH19} to validate properties through fuzzing.
If the code fails any of the tests or violates its invariants at any point, \systemname{} iteratively attempts 
to repair the code until the maximum number of tries is exceeded or valid code is produced. 
If the code passes all tests, \systemname{} uses that code for the function and caches validated code
in a database on disk for use in future runs.
If the developer modifies a function's specification, the LLM will regenerate its code from scratch.
The code can also optionally be spliced into place in the source code and the \systemname{} header removed entirely---
the developer may use this later on to insert production-ready code.

\subsection{Example}
We demonstrate the capabilities of \systemname{} by implementing a ``Maximum Increasing Subarrays'' function that
takes in a list $nums$ of integers and attempts to find the maximum length $k$ such that $nums$ contains two consecutive
subarrays of length $k$ with strictly increasing elements. This is Problem \#3350 from LeetCode~\cite{leetcode}.
Instead of using a dataset like HumanEval~\cite{chen2021codex} on which GPT-4o already performs with high
accuracy~\cite{4oHumanEval} and suffers from data leakage~\cite{gptmodels}, we evaluate \systemname{} on a LeetCode
problem to take advantage of the challenging problem set and private test suites.

Figure~\ref{fig:headers} shows the signature and \systemname{} header for the \texttt{maxIncSubarrays} function
(the LeetCode problem description has been condensed for brevity of exposition).
When we provided \systemname{} with only the description and not the \texttt{tests} in the full specification,
\systemname{} produces the code in Figure~\ref{fig:answer-bad} on its first attempt and immediately accepts it, only
checking if the generated code compiles and runs. However, it is incorrect. Figure~\ref{fig:answer-bad} highlights
two errors in red: the code incorrectly searches for a single subarray of increasing values rather than two consecutive
subarrays and contains an off-by-one error. This faulty code passes only 42\% (469 of 1,111) of the tests in
LeetCode's private test suite.

When we provided \systemname{} with the full header in Figure~\ref{fig:headers}, including the unit tests, it produced
the code in Figure~\ref{fig:answer-good} that passes all 1,111 private tests. Corrections from the previous attempt are highlighted in green.
In response to failed tests, \systemname{} iteratively refined its initial code twice to produce the validated version.

This function demonstrates how the addition of tests can greatly improve the quality of the generated code over using a
natural language prompt alone, and how \systemname{} can be used to
validate the code produced by the LLM. In this case, the developer only provides a handful of unit tests, and \systemname{}
ensures that the generated code passes those tests, and the resulting code then passes over a thousand unseen tests.
We anticipate that the use of property-based tests will further help to consistently yield a high level of quality.

\subsection{Future Work}
Our initial \systemname{} prototype has demonstrated its potential to enhance the effectiveness of LLM-based code generation, and 
we are planning a variety of extensions to further improve its capabilities:

\textit{Guided Code Generation:} Based on past work on program sketching~\cite{DBLP:conf/aplas/Solar-Lezama09,DBLP:conf/pldi/Solar-LezamaRBE05},
we plan to enable the programmer to give \systemname{} and the underlying LLM more context for code generation via
developer-provided partial function implementations or pseudocode outlines.

\textit{Run-time Testing:} The \systemname{} validation steps may not expose all errors, particularly for unusual scenarios
or unanticipated edge cases. To mitigate this issue, we intend to extend \systemname{} to check a generated function's correctness
at run time on the function's real inputs to gain further confidence in the code's correctness.

\textit{Maintaining Correctness:} Programs often contain multiple functions with inter-dependent behaviors. We plan to extend
\systemname{} to support specifying, or potentially inferring, relationships between functions. This will help \systemname{}
ensure that modifications to one function's specifications are reflected in the code of all related functions so that their relationships are preserved.

\textit{Performance:} In this paper, we have focused primarily on functional correctness. However, \systemname{} is
poised to handle other aspects of code quality, such as performance. In the future, we plan on expanding \systemname{}
to also support specifications of run time or memory requirements. \systemname{} will similarly convey these requirements
to the LLM and validate them against the generated code via either testing or run-time analysis.

%% file: conclusion.tex
\section{Conclusion}

In this paper, we propose \systemname{}, a Python-embedded DSL that lets
developers generate validated, high-quality code via an LLM assistant that
uses tests and specifications to guide code generation. With this abstraction,
developers can focus more on the behavior of their code rather than the
implementation details and trust that the code produced by the LLM is robust, reliable, and efficient.

The \systemname{} prototype is open-source and has been downloaded over 5,000 times.
It is available at \href{https://github.com/plasma-umass/pythoness}{https://github.com/plasma-umass/pythoness}.

%% file: main.bbl
\begin{thebibliography}{10}
\providecommand{\url}[1]{#1}
\csname url@samestyle\endcsname
\providecommand{\newblock}{\relax}
\providecommand{\bibinfo}[2]{#2}
\providecommand{\BIBentrySTDinterwordspacing}{\spaceskip=0pt\relax}
\providecommand{\BIBentryALTinterwordstretchfactor}{4}
\providecommand{\BIBentryALTinterwordspacing}{\spaceskip=\fontdimen2\font plus
\BIBentryALTinterwordstretchfactor\fontdimen3\font minus
  \fontdimen4\font\relax}
\providecommand{\BIBforeignlanguage}[2]{{%
\expandafter\ifx\csname l@#1\endcsname\relax
\typeout{** WARNING: IEEEtran.bst: No hyphenation pattern has been}%
\typeout{** loaded for the language `#1'. Using the pattern for}%
\typeout{** the default language instead.}%
\else
\language=\csname l@#1\endcsname
\fi
#2}}
\providecommand{\BIBdecl}{\relax}
\BIBdecl

\bibitem{github_copilot}
{GitHub}, ``{GitHub Copilot},'' \url{https://github.com/features/copilot},
  2024, accessed: 2024-10-20.

\bibitem{openai_chatgpt}
{OpenAI}, ``{ChatGPT},'' \url{https://openai.com/chatgpt/}, 2024, accessed:
  2024-10-20.

\bibitem{paradis2024doesaiimpactdevelopment}
\BIBentryALTinterwordspacing
E.~Paradis, K.~Grey, Q.~Madison, D.~Nam, A.~Macvean, V.~Meimand, N.~Zhang,
  B.~Ferrari-Church, and S.~Chandra, ``{How much does AI impact development
  speed? An enterprise-based randomized controlled trial},'' 2024. [Online].
  Available: \url{https://arxiv.org/abs/2410.12944}
\BIBentrySTDinterwordspacing

\bibitem{DBLP:conf/chi/Vaithilingam0G22}
P.~Vaithilingam, T.~Zhang, and E.~L. Glassman, ``Expectation vs. experience:
  Evaluating the usability of code generation tools powered by large language
  models,'' in \emph{{CHI} '22: {CHI} Conference on Human Factors in Computing
  Systems, Extended Abstracts}, 2022, pp. 332:1--332:7.

\bibitem{DBLP:journals/pacmpl/BarkeJP23}
S.~Barke, M.~B. James, and N.~Polikarpova, ``Grounded copilot: How programmers
  interact with code-generating models,'' \emph{Proc. {ACM} Program. Lang.},
  vol.~7, no. {OOPSLA1}, pp. 85--111, 2023.

\bibitem{chopra2023conversational}
B.~Chopra, A.~Singha, A.~Fariha, S.~Gulwani, C.~Parnin, A.~Tiwari, and A.~Z.
  Henley, ``{Conversational Challenges in AI-Powered Data Science: Obstacles,
  Needs, and Design Opportunities},'' \emph{arXiv preprint arXiv:2310.16164},
  2023.

\bibitem{kalliamvakou2022research}
E.~Kalliamvakou, ``{Research: Quantifying GitHub Copilot's Impact on Developer
  Productivity and Happiness},'' \emph{The GitHub Blog}, 2022.

\bibitem{liang2024usability}
J.~T. Liang, C.~Yang, and B.~A. Myers, ``{A Large-Scale Survey on the Usability
  of AI Programming Assistants: Successes and Challenges},'' in
  \emph{Proceedings of the 46th IEEE/ACM International Conference on Software
  Engineering}, 2024, pp. 1--13.

\bibitem{pythia}
\BIBentryALTinterwordspacing
(2024) Pythia. Accessed: 2024-11-16. [Online]. Available:
  \url{https://en.wikipedia.org/wiki/Pythia}
\BIBentrySTDinterwordspacing

\bibitem{leetcode}
\BIBentryALTinterwordspacing
(2024) {LeetCode}. Accessed: 2024-11-16. [Online]. Available:
  \url{https://leetcode.com/problemset/}
\BIBentrySTDinterwordspacing

\bibitem{DBLP:conf/nips/ZelikmanHPGH23}
\BIBentryALTinterwordspacing
E.~Zelikman, Q.~Huang, G.~Poesia, N.~D. Goodman, and N.~Haber, ``Parsel:
  Algorithmic reasoning with language models by composing decompositions,'' in
  \emph{Advances in Neural Information Processing Systems 36: Annual Conference
  on Neural Information Processing Systems 2023, NeurIPS 2023, New Orleans, LA,
  USA, December 10 - 16, 2023}, A.~Oh, T.~Naumann, A.~Globerson, K.~Saenko,
  M.~Hardt, and S.~Levine, Eds., 2023. [Online]. Available:
  \url{http://papers.nips.cc/paper\_files/paper/2023/hash/6445dd88ebb9a6a3afa0b126ad87fe41-Abstract-Conference.html}
\BIBentrySTDinterwordspacing

\bibitem{DBLP:conf/iclr/ChenZNZLLC23}
\BIBentryALTinterwordspacing
B.~Chen, F.~Zhang, A.~Nguyen, D.~Zan, Z.~Lin, J.~Lou, and W.~Chen, ``Codet:
  Code generation with generated tests,'' in \emph{The Eleventh International
  Conference on Learning Representations, {ICLR} 2023, Kigali, Rwanda, May 1-5,
  2023}.\hskip 1em plus 0.5em minus 0.4em\relax OpenReview.net, 2023. [Online].
  Available: \url{https://openreview.net/forum?id=ktrw68Cmu9c}
\BIBentrySTDinterwordspacing

\bibitem{DBLP:journals/sigsoft/FinkB97}
G.~Fink and M.~Bishop, ``{Property-Based Testing: A New Approach to Testing for
  Assurance},'' \emph{{ACM} {SIGSOFT} Softw. Eng. Notes}, vol.~22, no.~4, pp.
  74--80, 1997.

\bibitem{DBLP:journals/jossw/MaciverH19}
D.~Maciver and Z.~Hatfield{-}Dodds, ``{Hypothesis: A New Approach to
  Property-Based Testing},'' \emph{J. Open Source Softw.}, vol.~4, no.~43, p.
  1891, 2019.

\bibitem{chen2021codex}
OpenAI, ``{Evaluating Large Language Models Trained on Code},'' 2021.

\bibitem{4oHumanEval}
\BIBentryALTinterwordspacing
------. (2024) {Hello GPT-4o}. Accessed: 2024-11-16. [Online]. Available:
  \url{https://openai.com/index/hello-gpt-4o/}
\BIBentrySTDinterwordspacing

\bibitem{gptmodels}
\BIBentryALTinterwordspacing
------. (2024) Models. Accessed: 2024-11-16. [Online]. Available:
  \url{https://platform.openai.com/docs/models/gpt-4-turbo-and-gpt-4}
\BIBentrySTDinterwordspacing

\bibitem{DBLP:conf/aplas/Solar-Lezama09}
A.~Solar{-}Lezama, ``{The Sketching Approach to Program Synthesis},'' in
  \emph{Proceedings of the Asian Symposium on Programming Languages and
  Systems}, ser. Lecture Notes in Computer Science, vol. 5904.\hskip 1em plus
  0.5em minus 0.4em\relax Springer, 2009, pp. 4--13.

\bibitem{DBLP:conf/pldi/Solar-LezamaRBE05}
A.~Solar{-}Lezama, R.~M. Rabbah, R.~Bod{\'{\i}}k, and K.~Ebcioglu,
  ``{Programming by Sketching for Bit-Streaming Programs},'' in
  \emph{Proceedings of the {ACM} {SIGPLAN} 2005 Conference on Programming
  Language Design and Implementation}.\hskip 1em plus 0.5em minus 0.4em\relax
  {ACM}, 2005, pp. 281--294.

\end{thebibliography}
